\documentclass[12pt]{article}

\usepackage{latexsym,amssymb,epsfig}

\DeclareFontFamily{OT1}{rsfs10}{}
\DeclareFontShape{OT1}{rsfs10}{m}{n}{ <-> rsfs10 }{}
\DeclareMathAlphabet{\mathscript}{OT1}{rsfs10}{m}{n}

\newcommand{\eref}[1]{(\ref{#1})}

\newcommand{\fref}[1]{Figure~\ref{#1}}
\newcommand{\cref}[1]{Chapter~\ref{#1}}
\newcommand{\bcenter}{\begin{center}}
\newcommand{\ecenter}{\end{center}}
\newcommand{\beq}{\begin{equation}}
\newcommand{\eeq}{\end{equation}}
\newcommand{\bea}{\begin{eqnarray}}
\newcommand{\eea}{\end{eqnarray}}
\newcommand{\bean}{\begin{eqnarray*}}
\newcommand{\eean}{\end{eqnarray*}}
\newcommand{\ba}{\begin{array}}
\newcommand{\ea}{\end{array}}
\newcommand{\ben}{\begin{enumerate}}
\newcommand{\een}{\end{enumerate}}
\newcommand{\bi}{\begin{itemize}}
\newcommand{\ei}{\end{itemize}}
\newcommand{\bd}{\begin{description}}
\newcommand{\ed}{\end{description}}
\def\fnote#1#2{\begingroup\def\thefootnote{#1}\footnote{#2}
     \addtocounter{footnote}{-1}\endgroup}

\def\IF{\mathbb{F}}
\def\IZ{\mathbb{Z}}

\def\IP{\mathbb{P}}

\def\cN{{\mathcal N}}

\def\cE{{\mathcal E}}

\def\cO{{\mathcal O}}
\def\cC{{\mathcal C}}

\def\nn{\nonumber}

\def\rk{\mbox{rk}}

\def\av{\wedge^2 V}
\def\vv{V \otimes V^*}

\begin{document}

\begin{titlepage}

\vspace{-2cm}

\title{
   \hfill{\normalsize  UPR-1056-T} \\[1em]
   {\LARGE Moduli Dependent Spectra of Heterotic Compactifications}
\author{Ron Donagi$^1$, Yang-Hui He$^2$, Burt A.~Ovrut$^2$, 
	and Ren\'{e} Reinbacher$^3$
	\fnote{~}{donagi@math.upenn.edu;
	yanghe,	ovrut@physics.upenn.edu;
	rreinb@physics.rutgers.edu}\\[0.5cm]
   {\normalsize $^1$ 
	Department of Mathematics, University of Pennsylvania} \\
   {\normalsize $^2$
	Department of Physics, University of Pennsylvania} \\
   {\normalsize Philadelphia, PA 19104--6396, USA} \\
   {\normalsize $^3$
	Department of Physics and Astronomy, Rutgers University}\\
   {\normalsize Piscataway, NJ 08855-0849, USA}}
}
\date{}

\maketitle

\begin{abstract}
Explicit methods are presented for computing the cohomology of stable,
holomorphic vector bundles on elliptically fibered Calabi-Yau
threefolds. The complete particle spectrum of the low-energy,
four-dimensional theory is specified by the dimensions of
specific cohomology groups. The spectrum is shown to depend on the
choice of vector bundle moduli, jumping up from a generic minimal
result to attain many higher values on
subspaces of co-dimension one or higher in the moduli space. An explicit
example is presented within the context of a heterotic
vacuum corresponding to an $SU(5)$ GUT in four-dimensions.
\end{abstract}

\thispagestyle{empty}

\end{titlepage}

One approach to producing phenomenologically viable $N=1$
supersymmetric physics from strongly coupled $E_8 \times E_8$
heterotic string theory \cite{hw}
is to compactify the ten-dimensional
spacetime on a Calabi-Yau threefold $X$. Additionally, it is required
that the $E_8 \times E_8$ gauge connection $A$ satisfy the hermitian
Yang-Mills equations
\beq\label{YM}
F_{ab} = F_{\bar{a}\bar{b}} = g^{a\bar{b}}F_{\bar{b}a} = 0
\eeq
on $X$, where $F$ is the field strength of $A$. In this paper, we will
ignore the $E_8$ factor in the ``hidden'' sector, restricting the
subsequent discussion to the remaining $E_8$ gauge group. Hence, $A$
and $F$ will be Lie algebra valued over a single $E_8$. We call an
$E_8$ gauge connection satisfying \eref{YM} a holomorphic instanton on
$X$. Calabi-Yau threefolds are easily constructed. However, one must
also find a holomorphic instanton to completely specify the
vacuum. Here, one runs into serious technical difficulties, since no
explicit
solutions of \eref{YM} on a Calabi-Yau threefold are known. How,
then, can one proceed? It was shown in \cite{UY} and \cite{Don} that
$A$ is a holomorphic instanton if and only if it is a connection on a
stable, holomorphic vector bundle $V$ with structure group $G
\subseteq E_8$ on $X$. Therefore, finding a solution of \eref{YM} is
completely equivalent to specifying such a vector bundle. Happily,
general methods for their construction have been found 
\cite{Donagi,FMW,dlow,dow,dopw,rt,rt2,klemm,DI}.

Let us choose $X$ and $V$. We first note that the gauge group $H$ of
the low energy, four-dimensional theory is the commutant in $E_8$ of
the structure group $G$. 
Now consider the vector supermultiplet of the
ten-dimensional theory which transforms as the adjoint 248
representation of $E_8$. With respect to $G
\times H$, the 248 representation decomposes as
\beq\label{decomp}
248 \rightarrow (1, Ad(H)) \oplus \bigoplus_i (R_i, r_i) \ ,
\eeq
where $Ad(H)$ specifies the adjoint representation of $H$ and
$\{(R_i,r_i)\}$ are some set of representations of $G$ and $H$
respectively. This indicates that the low energy theory will contain
$N=1$ supermultiplets transforming in the $Ad(H)$ and $\{r_i\}$
representations of $H$. How many supermultiplets in each
representation will occur? We first note that the $Ad(H)$
representation will always be a unique vector supermultiplet. That is,
\beq\label{n-ad}
n_{Ad(H)} = 1 \ .
\eeq
All other representations will be realized as chiral supermultiplets
in the low energy theory. The multiplicity of superfields transforming
in the representation $r_i$, however, is much more difficult to
compute. It is given by the dimension of the space of zero modes of
the Dirac operator on $X$ transforming in the associated $R_i$
representation of $G$. If we denote by $V_{R_i}$ the vector bundle
constructed from $V$ whose fibers transform in the $R_i$
representation, then this space of zero modes is the cohomology group
$H^1(X, V_{R_i})$. It follows that the number of chiral
supermultiplets carrying the representation $r_i$ is given by
\beq\label{n-ri}
n_{r_i} = h^1(X, V_{R_i}) \ ,
\eeq
where lower case $h$ indicates the dimension. Therefore, to calculate
the spectrum of the low energy theory, one must compute the dimensions
of $H^1(X, V_{R_i})$ for each representation $R_i$ occurring in the
decomposition \eref{decomp}. For the special case when
\beq\label{TX}
V = TX \ ,
\eeq
the so-called ``standard'' embedding, these calculations are
relatively straight-forward, the results being related to the known
Betti numbers of the Calabi-Yau threefold \cite{GSW}. However,
for any stable, holomorphic vector bundle not satisfying \eref{TX},
the vast majority of such bundles, the computation of the spectrum is
a difficult problem. In this paper, we will present explicit
methods for calculating \eref{n-ri} in a wide class of
phenomenologically relevant heterotic string compactifications.

Before proceeding, however, we give a specific example of the above
remarks. Let us choose the structure group of $V$ to be $G =
SU(5)$. Then, the low energy gauge group is clearly $H = SU(5)$.
To distinguish the two different $SU(5)$ groups, we henceforth denote
them by $SU(5)_G$ and $SU(5)_H$ respectively. With respect to
$SU(5)_G \times SU(5)_H$, the adjoint representation
of $E_8$ decomposes as
\beq\label{decomp-5}
248 \rightarrow (1,24)  \oplus  (24,1) \oplus (5,\overline{10}) 
	\oplus (\overline{5},10) \oplus
	(10, 5) \oplus (\overline{10},\overline{5}) \ .
\eeq
The first term is $(1,Ad(SU(5)_H))$. Therefore, \eref{n-ad} implies
\beq\label{n24}
n_{24} = 1 \ .
\eeq
Now consider the remaining terms in \eref{decomp-5}. Noting that the
vector bundles corresponding to $R = 24, 5, \overline{5}, 10,
\overline{10}$ are
\beq
V_R = \vv, V, V^*, \av, \av^*
\eeq
respectively, it follows from \eref{decomp-5} and \eref{n-ri} that
\beq\label{n-vv}
n_1 = h^1(X, \vv)
\eeq
and
\bea\label{n-vvd}
&&n_{\overline{10}} = h^1(X,V), \quad n_{10} = h^1(X,V^*), \nn \\ 
&&n_{5} = h^1(X,\av), \quad n_{\overline{5}} = h^1(X, \av^*) \ .
\eea
Note that ${\rm Tr}(V \otimes V^*)$ can be ignored when computing
$h^1(X,\vv)$ on a Calabi-Yau threefold $X$.

Returning to the general decomposition \eref{decomp}, we must confront
the problem of how to compute the spectrum given in \eref{n-ri}. An
important insight is that the Atiyah-Singer index theorem will relate
the dimensions of different cohomology groups, thereby reducing the
amount of calculation. Using Serre duality, the stability of $V_{R_i}$
and the fact that $c_1(TX) = c_1(V_{R_i}) = 0$, the index theorem
gives
\beq\label{AS}
-h^1(X,V_{R_i})+h^1(X,V_{R_i}^*) = \frac12 \int_X c_3(V_{R_i}) \ ,
\eeq
for any representation $R_i$ of $G$. When $V_{R_i}$ is self-dual, that
is, $V_{R_i} = V_{R_i}^*$, \eref{AS} reduces to $0=0$ and no
information is obtained. However, a bundle associated with 
part of the quark/lepton spectrum is
not self-dual. For such quark/lepton bundles, denoting by $V_{R_i}$
those constructed from $V$ only, we find that
$c_3(V_{R_i}) = c_3(V)$. For example, consider the specific case
presented in \eref{n-vvd}. Note that neither $V$ nor $\av$ are
self-dual. Furthermore, one can easily show that $c_3(\av) = c_3(V)$.
If we impose the phenomenological
constraint that the number of quark/lepton families be three, then $V$
must be chosen so that
\beq\label{Ngen}
c_3(V) = 6
\eeq
and the index theorem relation \eref{AS} becomes
\beq\label{Ngen2}
-n_{r_i} + n_{\bar{r}_i} = 3 \ .
\eeq
In this paper, we will henceforth impose constraint
\eref{Ngen}. Therefore, we need compute only one of $n_{r_i}$ and
$n_{\bar{r}_i}$, the other following from the index relation
\eref{Ngen2}. Applying this to the concrete example discussed above,
we learn that
\beq\label{Ngen3}
-n_{\overline{10}} + n_{10} = 3, \qquad
-n_{5} + n_{\overline{5}} = 3 \ .
\eeq
Hence, in this case, it will be sufficient to compute $h^1(X, \vv)$, 
$h^1(X, V)$ and $h^1(X, \av^*)$, for example.

In order to explicitly compute the quantities $h^1(X, V_{R_i})$, we
must commit ourselves to a specific choice of both the Calabi-Yau
threefold $X$ and the stable, holomorphic vector bundle $V$. In this
paper, we will exploit the results of 
\cite{Donagi,FMW,dlow,dow,dopw,rt}
and choose $X$ to be
elliptically fibered over a base surface $B$, where $\pi: X
\rightarrow B$ is the projection map. To simplify our discussion, we
will take
\beq
B = \IF_r, \qquad r \in \IZ_{\ge 0}
\eeq
where $\IF_r$ are the Hirzebruch surfaces. It was shown in
\cite{Donagi,FMW,dlow} that a stable, holomorphic vector bundle 
can be constructed as follows. First, for specificity, choose
\beq
G = SU(n) \ .
\eeq
Then, consider a complex two-dimensional subspace $\cC_V$ of $X$ whose
homology class, which we also denote by $\cC_V$, is given by
\beq\label{CV}
\cC_V = n \sigma + \pi^* \eta \ ,
\eeq
where $\sigma$ is the image in $X$ of $\IF_r$, that is, a
zero section of the elliptic fiberation. Furthermore, $\eta$ is the
curve class
\beq\label{eta}
\eta = a S + b \cE \ , \quad a,b \in \IZ_{\ge 0}
\eeq
in $\IF_r$ with $S$ and $\cE$ the natural basis of $H_2(\IF_r,
\IZ)$. The subspace
$\cC_V$ is called a spectral cover. In addition, specify a line
bundle $\cN_V$ on $\cC_V$ by giving its first Chern class
\beq
\label{NV}
c_{1}(\cN_V)=n(\frac{1}{2}+\lambda)\sigma+(\frac{1}{2}-\lambda)
\pi^{*}\eta+(\frac{1}{2}+n\lambda)\pi^{*}c_{1}(T\IF_r) \ .
\eeq
Here, $\lambda = m \in \IZ$ if $n$ is even or 
$\lambda = m + \frac12$ if $n$ is
odd and
\beq
c_1(T\IF_r) = 2 S + (r+2) \cE
\eeq
is the first Chern class of $\IF_r$. It was shown in
\cite{Donagi,FMW,dlow} that one can construct a stable,
holomorphic vector bundle $V$ from $\cC_V$ and $\cN_V$ using a
Fourier-Mukai transformation
\beq\label{FM}
(\cC_V, \cN_V) \longleftrightarrow V \ .
\eeq
The details of this transformation are not important here. To simplify
the exposition in this paper, we will henceforth choose a specific
vacuum given by
\beq\label{eg}
r = 1, \quad
n =5, \quad
\eta = 12 S + 15 \cE, \quad
\lambda = \frac12 \ .
\eeq
We have shown that this vacuum satisfies the three fundamental
constraints required for the low energy theory
to be phenomenologically
relevant. That is, the theory is anomaly free, has three families of
quarks/leptons and is $N=1$ supersymmetric. The last two properties
are guaranteed by the fact that $V$ satisfies \eref{Ngen} and that $V$
is stable. Note that since $G$ and $H$ are $SU(5)_G$ and $SU(5)_H$
respectively, the low energy spectrum of this theory is given in
\eref{n24}, \eref{n-vv} and \eref{n-vvd} above. Furthermore, since
\eref{Ngen} is satisfied, the spectrum satisfies the index theorem
constraint in \eref{Ngen3}. To compute the spectrum, therefore, it
suffices to calculate 
$h^1(X, \vv)$, $h^1(X, V)$ and $h^1(X, \av^*)$.

We begin by considering $h^1(X, \vv)$. It is well-known that this
counts the moduli of the vector bundle $V$. It follows
\cite{transition,redumod} from \eref{FM} that
\beq\label{h1vv}
h^1(X, \vv) = (h^0(X, \cO_X(\cC_V)) - 1) + h^1(\cC_V, \cO_{\cC_V}).
\eeq
The first and second terms are the moduli of the spectral cover
$\cC_V$ and the line bundle $\cN_V$ respectively. Let us first compute 
$h^0(X, \cO_X(\cC_V))$. Noting that $\IF_1$ is itself a fiberation
over a base line $\IP^1$, and using \eref{CV}, \eref{eta} and
\eref{eg}, we find that
\bea\label{h0cV}
&&H^0(X, \cO_X(\cC_V)) \nn \\
&=&
\bigoplus_{i=3}^{15} H^0(\IP^1, \cO_{\IP^1}(i)) \oplus
\bigoplus_{i=1}^{9} H^0(\IP^1, \cO_{\IP^1}(i)) \oplus
\bigoplus_{i=0}^{6} H^0(\IP^1, \cO_{\IP^1}(i)) \nn \\
&& \quad
\oplus \bigoplus_{i=-1}^{3} H^0(\IP^1, \cO_{\IP^1}(i)) \oplus
\bigoplus_{i=-2}^{0} H^0(\IP^1, \cO_{\IP^1}(i)).
\eea
From the fact that $h^0(\IP^1, \cO_{\IP^1}(i)) = i+1$ for $i \ge
0$ and vanishes otherwise, it follows that $h^0(X, \cO_X(\cC_V)) =
223$. Now consider the second term $h^1(\cC_V, \cO_{\cC_V})$. 
Using an exact
sequence, the fact that $X$ is a Calabi-Yau threefold and Serre
duality, we can show that
\beq
h^1(\cC_V, \cO_{\cC_V}) =  h^1(X, \cO_X(\cC_V)) \ .
\eeq
Then, from \eref{CV}, \eref{eta}, \eref{eg} and the same techniques
used to compute \eref{h0cV}, one finds that $h^1(\cC_V, \cO_{\cC_V}) =
1$. Combining these two terms in \eref{h1vv} leads to the result
\beq
h^1(X,\vv) = 223.
\eeq
It then follows from \eref{n-vv} that
\beq
n_1 = 223
\eeq
is the number of vector bundle moduli in the low energy spectrum.

Let us now calculate $h^1(X,V)$. It is not too difficult to show that
\beq\label{h1XV}
h^1(X,V) = h^0(2S, L|_{2S}) \ , 
\eeq
where $2S$ arises as the intersection $\cC_V \cdot \sigma$,
$L$ is the line bundle
\beq
L = \cN_V \otimes \pi^* K_{\IF_1}
\eeq
and $K_{\IF_1}$ is the canonical bundle on $\IF_1$. Furthermore,
$L|_{2S}$ lies in the short exact sequence
\beq\label{seq}
0 \rightarrow \cO_{\IP^1}(-2) \rightarrow L|_{2S} \rightarrow
\cO_{\IP^1}(-3) \rightarrow 0 \ .
\eeq
Since neither $\cO_{\IP^1}(-2)$ nor $\cO_{\IP^1}(-3)$ has global
holomorphic sections, being of negative degree, it follows that
$L|_{2S}$ has no global sections. Therefore $h^0(2S, L|_{2S})$
vanishes and \eref{h1XV} implies that
\beq
h^1(X,V) = 0 \ .
\eeq
Expression \eref{n-vvd} and the index relation \eref{Ngen3} then give
\beq
n_{\overline{10}}  = 0, \quad  n_{10} = 3 \ .
\eeq

Finally, we compute $h^1(X, \av^*)$. Using a short exact sequence
similar to \eref{seq}, the 
Riemann-Roch theorem for a line bundle on the
curve $15(S + \cE)$ and several long exact cohomology sequences, we
find that
\beq\label{122}
h^1(X, \av^*) = 122 - \rk(M_2) \ .
\eeq
$M_2$ is a linear map defined by
\beq\label{M2}
H^1(X,W \otimes \cO_X(-\cC_V))_{180} 
\stackrel{M_2}{\longrightarrow} H^1(X,W)_{91}
\eeq
where
\beq
W = \pi^*\cO_B(14 S + 15 \cE)
\eeq
and $\cC_V$ is given by \eref{CV}, \eref{eta} and
\eref{eg}. The subscripts indicate that $H^1(X,W \otimes
\cO_X(-\cC_V))$ and $H^1(X,W)$ have dimension 180 and 91
respectively. Hence, $M_2$ can be represented as a $91 \times
180$ matrix. It follows from \eref{M2} that $M_2$ is parametrized by
elements of $H^0(X, \cO_X(\cC_V))$. Recall from the previous
discussion that this
is the 223 dimensional space of vector bundle moduli associated with
$\cC_V$.  A careful
analysis reveals that, of the 223 moduli, 139 parametrize $M_2$. The
matrix $M_2$ is composed of a large number of sub-matrices. The
non-vanishing sub-blocks break into two types, which we may write as
$m_{(4)i}$, $i=1, \ldots, 9$ and $m_{(6)j}$, $j=3,\ldots, 12$. Each
sub-matrix $m_{(4)i}$ is composed of $i+1$ vector bundle moduli, which
we denote by $\phi_p^{[(4)i]}$, $p=1,\ldots,i+1$. Similarly, each
$m_{(6)j}$ is composed of $j+1$ vector bundle moduli, denoted as 
$\phi_q^{[(6)j]}$ with $q=1, \ldots, j+1$. Note that the total number
of these moduli is 139, as stated above.

\begin{figure}
\centerline{\psfig{figure=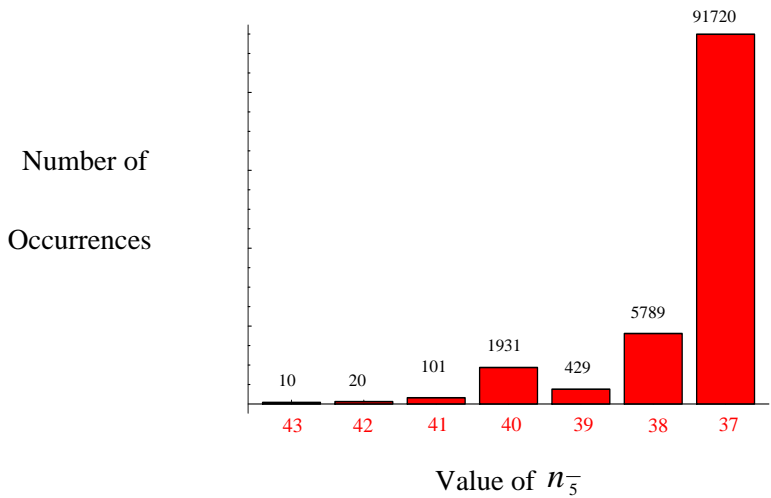,width=4in}}
\caption{In 100,000 random integer initializations of the matrix
$M_2$, the numbers of occurrences of the various values of
$n_{\overline{5}}$
are plotted. We see that the generic value 37 dominates by far. 
\label{f:random}}
\end{figure}

It follows from \eref{122} that to compute $h^1(X, \av^*)$, one must
calculate the rank of $M_2$. However, $M_2$ depends on 139 vector
bundle moduli. This opens the possibility that $\rk(M_2)$ depends on
where in moduli space it is evaluated. To explore this, we begin by
randomly selecting the values of all moduli, assuming that each is
non-zero. We then numerically compute the rank of $M_2$. Using
\eref{122}, we obtain $h^1(X, \av^*)$ and, via \eref{n-vvd}, a value
for $n_{\overline{5}}$. 
The results of a numerical calculation involving 100,000
random initializations are shown in \fref{f:random}. The moduli were
initialized to be positive integers in the range $[1,3]$. 
The horizontal axis indicates the values of $n_{\overline{5}}$
found in the survey, while the vertical axis gives the number of
occurrences. We see that the value 37 by far dominates over any other
possibilities. It follows that at generic points in moduli space
\beq\label{37}
n_{\bar{5}} = 37 \ .
\eeq

\begin{figure}
\centerline{\input{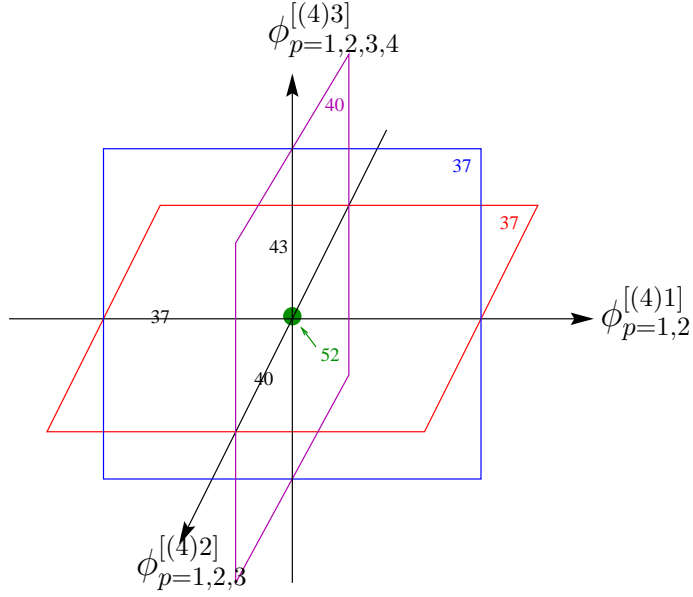}}
\caption{
A subspace of moduli space spanned
by $\phi^{[(4)1]}_{p=1,2}$, $\phi^{[(4)2]}_{q=1,2,3}$ and 
$\phi^{[(4)3]}_{r=1,2,3,4}$. Generically, in the bulk, 
$n_{\bar{5}} = 37$,
its minimal value. As we restrict to various planes and
intersections thereof, we are confining ourselves to special
sub-spaces of co-dimension one or higher. In these subspaces, the
value of $n_{\bar{5}}$ can increase dramatically.
\label{f:specjump}}
\end{figure}

Importantly, however, we notice that there are isolated
initializations of the moduli for which $n_{\bar{5}}$ 
jumps to values larger
than 37. This phenomenon is clearly seen in \fref{f:random} where
$n_{\bar{5}}$ 
is shown to attain all integer values between 38 and 43, in addition
to its generic value of 37. As we increase the number of
initializations, we expect to see even larger values for
$n_{\bar{5}}$. To test
this assertion, we now take a more analytic approach to the phenomenon
of the jumping of $n_{\bar{5}}$. Let us leave the sub-matrices $m_{(6)j}$
untouched and proceed to consecutively set the sub-blocks $m_{(4)i}$ to
zero. To begin, set $m_{(4)1}$ to zero by restricting the moduli to
the co-dimension two subspace defined by $\phi_1^{[(4)1]} =
\phi_2^{[(4)1]} = 0$. Now compute the rank of $M_2$ and, hence,
$n_{\bar{5}}$
numerically, initializing the remaining moduli to have random, but
non-zero, values. Generically, we find
\beq
n_{\bar{5}} = 40.
\eeq
Continuing in this way, setting the remaining $m_{(4)i}$ blocks to
zero one by one, we find additional values of $n_{\bar{5}}$ given by
\beq
n_{\bar{5}} = 43, 52, 61, 69, 76, 82, 87, 94 \ .
\eeq
Note that, as conjectured, values of $n_{\bar{5}}$ 
much larger than 43 are
obtained. As a graphic example of this phenomenon,
we show in \fref{f:specjump} a nine-dimensional region 
of the vector bundle moduli space discussed in
the previous analysis. Note that for a generic point in this space,
$n_{\bar{5}}=37$. 
However, on various sub-planes of co-dimension one or higher
$n_{\bar{5}}$ jumps, taking the values $n_{\bar{5}}=37,40,43$ and
$52$.  Various
analytic and numerical methods lead us to expect that $n_{\bar{5}}$ 
will, in fact, attain all integer values in the range
\beq\label{5-bars}
n_{\bar{5}} \in [37, 94] \ ,
\eeq
where, generically, $n_{\bar{5}} = 37$.
The index theorem relation \eref{Ngen3} tells us that
\beq
n_{5} = n_{\bar{5}} - 3 \ .
\eeq
It follows from \eref{37} that, generically,
\beq\label{34}
n_{5} = 34 \ .
\eeq
However, from \eref{34} we expect $n_{5}$ to attain all integer
values in the range
\beq\label{5-s}
n_{5} \in [34, 91] \ .
\eeq
This completes the evaluation of the low energy spectrum of the
heterotic vacuum specified in \eref{eg}. The number of 5 and $\bar{5}$
multiplets given in \eref{5-bars} and \eref{5-s} respectively is
rather large. This is entirely due to the explicit example we have
chosen. Much smaller generic values of 5 and $\bar{5}$ can be found in
other heterotic vacua.

Although we have computed the spectrum within the context of a
specific $SU(5)$ GUT, the techniques we have introduced, and the
phenomenon of the moduli dependence of the spectrum, are completely
general and will apply to any heterotic vacuum. 
In the standard embedding, the spectrum is related to the Hodge
numbers of the Calabi-Yau threefold \cite{GSW}. These, in turn, are
completely fixed by the Betti numbers, which are topological
invariants. Hence, the spectrum is moduli independent. However, for
any other stable, holomorphic vector bundle the spectrum is given by
$h^1(X, V_{R_i})$. Since $V$ has moduli, the associated spectrum is,
in general, dependent on these moduli.
A detailed analysis of
our method, as well as a discussion of the requisite mathematics, will
be presented elsewhere \cite{thing}.

\section*{Acknowlegements}
We are grateful to V.~Braun, E.~Buchbinder and 
T.~Pantev for many insightful comments and conversations.
This Research was supported in part by
the Dept.~of Physics and the Maths/Physics Research Group
at the University of Pennsylvania
under cooperative research agreement \#DE-FG02-95ER40893
with the U.~S.~Department of Energy, an NSF Focused Research Grant
DMS0139799 for ``The Geometry of Superstrings,'' and an
NSF grant DMS 0104354. R.~R.~is also supported
by the Department of Physics and Astronomy of Rutgers University under
grant DOE-DE-FG02-96ER40959.

\bibliographystyle{JHEP}

\end{document}